\documentstyle[aps,twocolumn]{revtex}
\begin{document}
\noindent {\large\bf{Comment on `` Theory of Diluted Magnetic
Semiconductor Ferromagnetism ''}} \vspace*{0.5cm}

In a recent Letter\cite{Konig}, a theory of carrier-induced
ferromagnetism in diluted magnetic semiconductors (DMS) is
proposed. By using their self-consistent spin-wave (SCSW)
approximation, the authors show a non-monotonic dependence of
critical temperatures $T_c$ on the itinerant-carrier density in
agreement with experiment. Here we emphasize that their SCSW
theory is {\it a priori} unjustified and will lead to incorrect
results at low temperatures and near $T_c$. Thus we suggest
another SCSW approximation to remedy these flaws.

By taking the Ising limit for the exchange coupling between
magnetic ions and itinerant carriers, such that the spin-wave
spectrum $\Omega_p$ is independent of momentum $\vec p$, one can
obtain an expression for the thermal average of the impurity-spin
density\cite{Konig}
\begin{equation}
   \langle S^z \rangle = {1\over V}\sum_{|\vec p| < p_c} \left\{ S - n(\Omega_p)
 + (2S+1) n\right[ (2S+1)\Omega_p \left] \right\},
\label{SelfConsistent1}
\end{equation}
where $n(\omega)$ is the Bose function, and $p_c$ is a Debye
cutoff. ($p_c^3 = 6 \pi^2 c$, $c$ is the magnetic ion density.)
The SCSW approximation used in Ref.\cite{Konig} consists of
extending the above formula (which is derived under the Ising
limit) to the isotropic case simply by substituting the $\Omega_p$
in the isotropic case (now $\Omega_p$ is $\vec p-$dependent) into
Eq.~(\ref{SelfConsistent1}). Thus their theory can be considered
phenomenological, and its validity is not guaranteed. For example,
as mentioned by themselves, when $T \rightarrow 0$, the third term
in Eq.~(\ref{SelfConsistent1}) is not negligible as compared with
the second term, consequently the prefactor of the characteristic
$T^{3/2}$ law for the localized-ions magnetization differs from
the correct value of the linearized spin-wave theory
(LSWT)\cite{Auerbach94}. Moreover, near $T_c$, where both $\langle
S^z \rangle$ and the itinerant-carrier spin density
$n^{*}=(n_\downarrow - n_\uparrow )/2$ approach zero, one can show
that Eq.~(\ref{SelfConsistent1}) leads to the following expression
for $T_c$,
\begin{equation}
   k_B T_c = {S(S+1)\over 3} \lim_{\langle S^z \rangle, n^{*}
   \rightarrow 0} {1\over V}\sum_{|\vec p| < p_c} {\Omega_p \over \langle S^z
   \rangle}.
\label{Tc1}
\end{equation}
Notice that, although the low-energy spin-wave excitations do exist
in the present system, this expression always predicts a nonzero
$T_c$, even for the one-dimensional (1D) and two-dimensional (2D)
cases! This is inconsistent with the Mermin-Wangner
theorem\cite{Auerbach94}, and implies that
Eq.~(\ref{SelfConsistent1}) does not properly capture the whole
effect of the spin fluctuations.

To find another SCSW theory without these flaws, we notice that,
after coarse graining as being done in Ref.~\cite{Konig}, the
Hamiltonian of DMS and the Kondo lattice model (KLM) are
approximately equivalent. Therefore, by using the
equation-of-motion approach under the Tyablikov decoupling
scheme\cite{Zubarev} and calculating the Green finction
$\langle\langle S_i^+ ; (S_j^-)^n (S_j^+)^{n-1} \rangle\rangle$
for the KLM, we obtain another expression for $\langle S^z
\rangle$\cite{GF},
\begin{equation}
   \langle S^z \rangle = c S- c \Phi +
    {(2S+1) c \over
    \left[ \left( 1+\Phi \right) /\Phi \right]^{2S+1} - 1},
\label{SelfConsistent2}
\end{equation}
where the value of $\langle S^z \rangle$ in the KLM is reduced by
a factor of $c$ due to coarse graining and $\Phi=(1/cV)\sum_{|\vec
p| < p_c} n(\Omega_p)$. As simple justification of
Eq.~(\ref{SelfConsistent2}), one can check two limiting cases: (i)
when $T \approx 0$ such that $n(\Omega_p)$ and therefore $\Phi$
are vanishingly small, the last term in
Eq.~(\ref{SelfConsistent2}) can be dropped, and then
Eq.~(\ref{SelfConsistent2}) reduces to the prediction of the LSWT;
(ii) by taking the Ising limit where $\Omega_p$ is $\vec
p-$independent, Eq.~(\ref{SelfConsistent1}) is recovered as it
should be. Moreover, our theory gives another expression for
$T_c$,
\begin{equation}
   k_B T_c =\frac{ S(S+1) / 3 }
   { \lim_{\langle S^z \rangle, n^{*} \rightarrow 0} (1/V) \sum_{|\vec p| < p_c}
   \langle S^z \rangle / \Omega_p c^2 }.
\label{Tc2}
\end{equation}
The above formula gives $T_c=0$ both for the 1D and 2D cases due
to the fact that $\Omega_p \propto p^2$ as $p \rightarrow 0$ and
therefore the integration over $\vec p$ diverges. Based on these
discussions, we argue that a reasonable SCSW theory should use
Eq.~(\ref{SelfConsistent2}) for $\langle S^z \rangle$, rather than
Eq.~(\ref{SelfConsistent1}). It should be noted that, because the
long-wavelength part of the spin waves plays a more important role
in our method as compared to theirs (cf. Eqs.~(\ref{Tc1}) and
(\ref{Tc2})), the physics near $T_c$ implied by using
Eqs.~(\ref{SelfConsistent1}) and (\ref{SelfConsistent2}) differ
qualitatively\cite{Pajda}.

\vspace*{0.5cm} \noindent Min-Fong Yang,$^1$ Shih-Jye Sun,$^2$ and
Ming-Che Chang$^3$

$^1$ Department of Physics,\\ \hspace*{0.51cm} Tunghai University,
Taichung, Taiwan,

$^2$ Department of Electronics,\\ \hspace*{0.51cm} Kao Yuan Institute
of Technology,\\ \hspace*{0.51cm} Kaohsiung, Taiwan,

$^3$ Department of Physics,\\ \hspace*{0.51cm} National Taiwan Normal
University, Taipei, Taiwan

\vspace*{0.5cm}

\noindent Received {\today}

\\

\end{document}